\documentclass[pra,aps,twocolumn,superscriptaddress]{revtex4}

\usepackage{amsmath}
\usepackage{bm}
\usepackage{graphicx}

\begin{document}

\title{Crossover between Kelvin-Helmholtz and counter-superflow
	instabilities in two-component Bose-Einstein condensates}

\author{Naoya Suzuki}
\affiliation{Department of Engineering Science, University of
Electro-Communications, Tokyo 182-8585, Japan}
\author{Hiromitsu Takeuchi}
\affiliation{Department of Physics, Osaka City University, Sumiyoshi-ku,
Osaka 558-8585, Japan}
\author{Kenichi Kasamatsu}
\affiliation{Department of Physics, Kinki University, Higashi-Osaka, Osaka
577-8502, Japan}
\author{Makoto Tsubota}
\affiliation{Department of Physics, Osaka City University, Sumiyoshi-ku,
Osaka 558-8585, Japan}
\author{Hiroki Saito}
\affiliation{Department of Engineering Science, University of
Electro-Communications, Tokyo 182-8585, Japan}

\date{\today}

\begin{abstract}
Dynamical instabilities at the interface between two Bose--Einstein
condensates that are moving relative to each other are investigated using
mean-field and Bogoliubov analyses.
Kelvin--Helmholtz instability is dominant when the interface thickness
is much smaller than the wavelength of the unstable interface mode,
whereas the counter-superflow instability becomes dominant in the opposite
case.
These instabilities emerge not only in an immiscible system but also in a
miscible system where an interface is produced by external potential.
Dynamics caused by these instabilities are numerically demonstrated in
rotating trapped condensates.
\end{abstract}


\maketitle

\section{Introduction}

When wind blows over a water surface, the relative motion between the
water and the air generates instabilities at the interface;
these instabilities in turn produce waves.
Such interfacial instabilities between two fluids that are moving relative
to each other are referred to as Kelvin--Helmholtz instabilities
(KHIs)~\cite{Helm,Kelvin,Lamb} and they are found throughout nature.
A system of superfluids is an ideal testing ground for KHIs because of
the absence of viscosity.
The Helsinki group experimentally realized shear flow between the A and
B phases of superfluid $^3 {\rm He}$ in a rotating cryostat.
They observed vortices penetrating from the A phase into the B phase due
to KHIs~\cite{Blaa,Volovik}.

Recently, KHIs and nonlinear dynamics of quantized vortices have been
investigated for a gaseous two-component Bose--Einstein condensate (BEC)
with relative velocity~\cite{Takeuchi}.
In Ref.~\cite{Takeuchi}, it is assumed that the two components are
strongly segregated and that the interface thickness is much smaller
than the wavelengths of unstable interface modes.
If, however, the two components are weakly segregated, they strongly
overlap with each other at the interface, at which the two components
coexist with a relative velocity.
Law {\it et al.}~\cite{Law} showed that two miscible BECs moving through
each other exhibit a dynamical instability, which we refer to as a
counter-superflow instability (CSI).
The dynamics generated by CSIs have recently been experimentally
observed~\cite{Hamner}.
We therefore expect that KHIs compete with CSIs as the thickness of
the interface is increased and that CSIs become important in the
weakly segregated regime.

In the present paper, we investigate both KHI and CSI on an equal footing
in a phase-separated two-component BEC with relative velocity.
We show that KHIs and CSIs are respectively generated for thin and thick
interfaces;
the interface thickness is controlled by the repulsive interaction between
the two components and external forces.
The numerically obtained stability boundaries in the parameter space are
well described by the KHI in the thin interface limit and by the CSI in
the uniform overlap limit.
We propose various experimental situations for observing these
instabilities in trapped systems.

This paper is organized as follows.
Section~\ref{s:ideal} reviews KHIs and CSIs in BECs, and performs the
mean-field and Bogoliubov analyses for a system with a flat interface.
Sections~\ref{s:quasi} and \ref{s:3D} propose experimental systems for
generating these instabilities in pancake-shaped and cigar-shaped traps,
respectively, and numerically demonstrate the nonlinear dynamics caused by
these instabilities.
Section~\ref{s:conclusion} presents the conclusions of this study.

\section{Instabilities at an ideal interface}
\label{s:ideal}

\subsection{Kelvin--Helmholtz instability}
\label{s:KHI}

We first formulate a KHI in a strongly segregated two-component BEC
with relative velocity between its two components~\cite{Takeuchi2}.
In the mean-field theory, the two-component BEC is described by the
macroscopic wave functions $\psi_1$ and $\psi_2$.
The Lagrangian for the system is given by
\begin{equation} \label{L}
L = \int d\bm{r} \left( P_1 + P_2 - g_{12} |\psi_1|^2 |\psi_2|^2
\right),
\end{equation}
where
\begin{equation}
P_j = i \hbar \psi_j^* \frac{\partial \psi_j}{\partial t} +
\frac{\hbar^2}{2 m_j} \psi_j^* \nabla^2 \psi_j - U_j |\psi_j|^2 -
\frac{g_{jj}}{2} |\psi_j|^4
\end{equation}
with $m_j$ and $U_j$ being the atomic mass and the external potential
of the $j$th component, respectively.
The inter- and intra-component interaction parameters have the form,
\begin{equation}
g_{jj'} = 2\pi \hbar^2 a_{jj'} (m_j^{-1} + m_{j'}^{-1}),
\end{equation}
where $a_{jj'}$ is the s-wave scattering length between the $j$th and
$j'$th components.
The interaction parameters are assumed to satisfy the phase-separation
condition~\cite{Pethick},
\begin{equation} \label{immiscible}
g_{11} g_{22} < g_{12}^2.
\end{equation}

We assume that components 1 and 2 are respectively located in $y \lesssim
0$ and $y \gtrsim 0$ and that the interface between the two components is
located near the $y = 0$ plane.
In this subsection, we neglect the interface thickness and introduce an
interface tension coefficient $\alpha$~\cite{Ao,Barankov,Schae}, which
originates from the excess energy at the interface.
The Lagrangian can thus be rewritten as
\begin{equation}
L = \int dx dz \left( \int_{-\infty}^{\eta} dy P_1 +
\int_{\eta}^{\infty} dy P_2 \right) - \alpha S,
\end{equation}
where $y = \eta(x, z, t)$ is the position of the interface and
\begin{eqnarray}
S & = & \int dx dz \left[ 1 + \left( \frac{\partial \eta}{\partial x}
\right)^2 + \left( \frac{\partial \eta}{\partial z} \right)^2
\right]^{1/2} \nonumber \\
& \simeq & \int dx dz \left[ 1 + \frac{1}{2} \left( \frac{\partial
\eta}{\partial x} \right)^2 + \frac{1}{2} \left( \frac{\partial
\eta}{\partial z} \right)^2 \right]
\end{eqnarray}
is the area of the interface.
Taking the functional derivative of the action $\int dt L$ with respect
to $\eta(x, z, t)$ and setting it to zero, we obtain
\begin{equation} \label{Ber}
P_1(y = \eta) - P_2(y = \eta) + \alpha \left( \frac{\partial^2
\eta}{\partial x^2} + \frac{\partial^2 \eta}{\partial z^2} \right) = 0,
\end{equation}
which corresponds to the Bernoulli equation in hydrodynamics.

We consider a stationary state in which the $j$th component flows in the
$x$ direction with a velocity $V_j$ as
\begin{equation} \label{Psi}
\Psi_j = \sqrt{n_j(y)} \exp\left[\frac{i}{\hbar} \left( -\mu_j t
+ m_j V_j x \right) \right],
\end{equation}
where $\mu_j$ is the chemical potential for the $j$th component.
The potential $U_j(y)$ and the density distribution $n_j(y)$ are assumed
to depend only on $y$, where $n_1 = 0$ for $y > \eta$ and $n_2 = 0$ for $y
< \eta$ are satisfied.
Substituting Eq.~(\ref{Psi}) into Eq.~(\ref{Ber}) with $\eta = 0$ gives
the equilibrium condition for the pressure, $g_{11} n_1^2 / 2 = g_{22}
n_2^2 / 2$.

We assume that the system is approximately incompressible.
A small deviation from the stationary state in this case is described as
\begin{eqnarray} \label{dpsi}
\psi_j & = & \Psi_j \exp\left[ i A_j e^{-(-1)^j k y} \cos(k x - \omega
t) \right], \\
\eta & = & a \sin(k x - \omega t),
\label{eta}
\end{eqnarray}
where $A_j$ and $a$ are infinitesimal parameters.
From the kinematic boundary condition, the interface velocity in
the $y$ direction $(\partial / \partial t + V_j \partial / \partial x)
\eta$ must be equal to $\hbar / (i m_j n_j) \psi_j^* \partial \psi_j /
\partial y |_{y = \eta}$, giving
\begin{equation} \label{aArel}
-(-1)^j \frac{\hbar}{m_j} A_j k e^{-(-1)^j k \eta} = (V_j k - \omega) a.
\end{equation}
Substituting Eqs.~(\ref{dpsi})-(\ref{aArel}) into Eq.~(\ref{Ber}) and
neglecting second and higher orders of $A_j$ and $a$, we obtain
\begin{equation} \label{omega}
\frac{\rho_1}{k} (\omega - V_1 k)^2 - f_1 n_{{\rm s} 1} =
 -\frac{\rho_2}{k} (\omega - V_2 k)^2 - f_2 n_{{\rm s} 2} + \alpha k^2,
\end{equation}
where $n_{{\rm s} 1} = n_1(\eta + 0_-)$, $n_{{\rm s} 2} = n_2(\eta +
0_+)$, $\rho_j = m_j n_{{\rm s} j}$, and $f_j = U_j'(\eta)$.
Equation~(\ref{omega}) gives the dispersion relation,
\begin{equation} \label{KHId}
\omega = \frac{(\rho_1 V_1 + \rho_2 V_2) k}{\rho_1 + \rho_2} \pm
 \sqrt{-\frac{\rho_1 \rho_2 (V_1 - V_2)^2 k^2}{(\rho_1 + \rho_2)^2} +
 \frac{F k + \alpha k^3}{\rho_1 + \rho_2}},
\end{equation}
where $F = n_{{\rm s} 1} f_1 - n_{{\rm s} 2} f_2$.

We note that Eq.~(\ref{KHId}) has the same form as the dispersion
relation for the KHI in classical incompressible and inviscid
fluids~\cite{Lamb}.
When $F > 0$ and
\begin{equation} \label{Vcr}
(V_1 - V_2)^4 \geq 4 \alpha F \frac{(\rho_1 + \rho_2)^2}{\rho_1^2
		\rho_2^2} \equiv V_{\rm cr}^4,
\end{equation}
a dynamical instability arises for $k_- < k < k_+$, where
\begin{equation}
k_\pm = \frac{\rho_1 \rho_2 (V_1 - V_2)^2}{2\alpha (\rho_1 + \rho_2)}
\left[ 1 \pm \sqrt{1 - \frac{V_{\rm cr}^4}{(V_1 - V_2)^4}} \right].
\end{equation}
The most unstable wave number is given by
\begin{equation} \label{mu}
k = \frac{\rho_1 \rho_2 (V_1 - V_2)^2}{3\alpha (\rho_1 + \rho_2)}
\left[ 1 + \sqrt{1 - \frac{3 V_{\rm cr}^4}{4 (V_1 - V_2)^4}} \right].
\end{equation}
For $|V_1 - V_2| < V_{\rm cr}$, the system is dynamically stable.
When $F = 0$, $V_{\rm cr}$ vanishes and the system is dynamically unstable
for $|V_1 - V_2| > 0$.
In this case, the range of unstable wave numbers is given by
\begin{equation} \label{KHIrange}
0 < k < \frac{\rho_1 \rho_2 (V_1 - V_2)^2}{\alpha (\rho_1 + \rho_2)}.
\end{equation}
From Eq.~(\ref{KHId}), we find that the group velocity $d \omega / d k$
diverges for $k \rightarrow 0$ and $F > 0$.
This unphysical behavior originates from the incompressibility
approximation in Eq.~(\ref{dpsi}).
For $F < 0$, the system is always dynamically unstable even when $V_1 -
V_2 = 0$, which is referred to as a Rayleigh--Taylor
instability~\cite{Rayleigh,Taylor,Sasaki}.

\subsection{Counter-superflow instability}
\label{s:counter}

We review the derivation of the Bogoliubov spectrum for a system of a
uniform two-component BEC with a relative velocity~\cite{Law}.
The functional derivative of $\int dt L$, where $L$ is the Lagrangian in
Eq.~(\ref{L}), with respect to $\psi_j^*$ gives the Gross-Pitaevskii
(GP) equation $(j \neq j')$,
\begin{equation} \label{GP}
i \hbar \frac{\partial \psi_j}{\partial t} = -\frac{\hbar^2}{2 m_j}
\nabla^2 \psi_j + U_j \psi_j + g_{jj} |\psi_j|^2 \psi_j + g_{jj'}
|\psi_{j'}|^2 \psi_j.
\end{equation}
We assume $U_j = 0$ and the miscible condition,
\begin{equation} \label{miscible}
g_{11} g_{22} > g_{12}^2,
\end{equation}
in this subsection.
We consider a small excitation $\phi_j$ above a uniform state with a
velocity $\bm{V}_j = \hbar \bm{K}_j / m_j$ as
\begin{equation} \label{phi}
\psi_j = \left( \sqrt{n_j} + \phi_j \right) e^{-i \mu_j t / \hbar + i
 \bm{K}_j \cdot \bm{r}},
\end{equation}
where $\mu_j = g_{jj} n_j + g_{jj'} n_{j'} + \hbar^2 K_j^2 / (2m_j)$.
Substituting Eq.~(\ref{phi}) into Eq.~(\ref{GP}) and taking the first
order of $\phi_j$, we obtain $(j \neq j')$
\begin{eqnarray} \label{bogo}
i \hbar \frac{\partial \phi_j}{\partial t} & = & \left[
-\frac{\hbar^2}{2m_j} \left( \nabla + i \bm{K}_j \right)^2 - \mu_j + 2
g_{jj} n_j + g_{jj'} n_{j'} \right] \phi_j \nonumber \\
& & + g_{jj} n_j \phi_j^* +
g_{jj'} \sqrt{n_j n_{j'}} \left( \phi_{j'} + \phi_{j'}^* \right).
\end{eqnarray}
We expand the small excitation as
\begin{equation}
\phi_j = u_{j \bm{k}} e^{i \bm{k} \cdot \bm{r} - i \omega t}
- v_{j \bm{k}}^* e^{-i \bm{k} \cdot \bm{r} + i \omega t},
\end{equation}
and substitute it into Eq.~(\ref{bogo}), giving $(j \neq j')$
\begin{widetext}
\begin{subequations} \label{bogo2}
\begin{eqnarray}
\left[ \frac{\hbar^2}{2m_j} \left( k^2 + 2 \bm{k} \cdot \bm{K}_j \right)
+ g_{jj} n_j \right] u_{j \bm{k}} - g_{jj} n_j v_{j \bm{k}} 
+ g_{jj'}
\sqrt{n_j n_{j'}} \left( u_{j' \bm{k}} - v_{j' \bm{k}} \right) & = & \hbar
\omega u_{j \bm{k}}, \\
\left[ \frac{\hbar^2}{2m_j} \left( k^2 - 2 \bm{k} \cdot \bm{K}_j \right)
+ g_{jj} n_j \right] v_{j \bm{k}} - g_{jj} n_j u_{j \bm{k}} 
- g_{jj'} \sqrt{n_j n_{j'}} \left( u_{j' \bm{k}} - v_{j' \bm{k}}
\right) & = & -\hbar \omega v_{j \bm{k}}.
\end{eqnarray}
\end{subequations}
Diagonalizing the eigenvalue equation (\ref{bogo2}), we obtain the
Bogoliubov excitation spectrum.

For simplicity, we assume that $m_1 = m_2 \equiv m$ and $g_{11} n_1 =
g_{22} n_2 \equiv u$.
Then, the eigenvalue of Eq.~(\ref{bogo2}) has a simple form
\begin{equation} \label{counter}
\hbar \omega = \frac{\hbar}{2} (\bm{V}_1 + \bm{V}_2) \cdot \bm{k} \pm
 \left[ \varepsilon_0^2 + \varepsilon_{\rm r}^2 + 2 \varepsilon_0 u \pm
	2 \left( \varepsilon_0^2 \varepsilon_{\rm r}^2 + 2 \varepsilon_0
		 \varepsilon_{\rm r}^2 u + \varepsilon_0^2 u_{12}^2
\right)^{1/2} \right]^{1/2},
\end{equation}
where $\varepsilon_0 = \hbar^2 k^2 / (2 m)$, $\varepsilon_{\rm r} =
\hbar \bm{k} \cdot (\bm{V}_1 - \bm{V}_2) / 2$, and $u_{12} = g_{12} (n_1
n_2)^{1/2}$.
The expression in the square brackets in Eq.~(\ref{counter}) for the
negative sign becomes negative for
\begin{equation} \label{krange}
{\rm max} [k_{\rm r}^2 \cos^2 \chi - 4m (u + u_{12}) / \hbar^2, 0] <
 k^2 < {\rm max} [k_{\rm r}^2 \cos^2 \chi - 4m (u - u_{12}) / \hbar^2,
 0],
\end{equation}
\end{widetext}
where the function max yields the maximum value of the arguments,
$\chi$ is the angle between $\bm{V}_1 - \bm{V}_2$ and $\bm{k}$, and
$k_{\rm r} = m |\bm{V}_1 - \bm{V}_2| / \hbar$.
From Eq.~(\ref{krange}), the system is dynamically stable when $k_{\rm
r}^2 < 4 m (u - u_{12}) / \hbar^2$.
Minimizing the expression in the square bracket of Eq.~(\ref{counter})
with respect to $k$ and $\chi$, we find $\cos^2\chi = 1$;
thus, the most unstable wave vector is parallel to the relative velocity.

\subsection{Crossover between Kelvin--Helmholtz and counter-superflow
instabilities}

In Sec.~\ref{s:counter}, we showed that a CSI emerges when the two
components overlap with a relative velocity.
We expect that a similar situation arises when the interface between the
two components is sufficiently thick in the system discussed in
Sec.~\ref{s:KHI}, since in the interface region the two components
overlap considerably and have a relative velocity.
In this subsection, we show that a CSI emerges when the interface
thickness is much larger than the characteristic wavelength.
We also show that similar behavior is observed in miscible condensates
separated by a potential gradient.

\begin{figure*}[t]
\includegraphics[width=17cm]{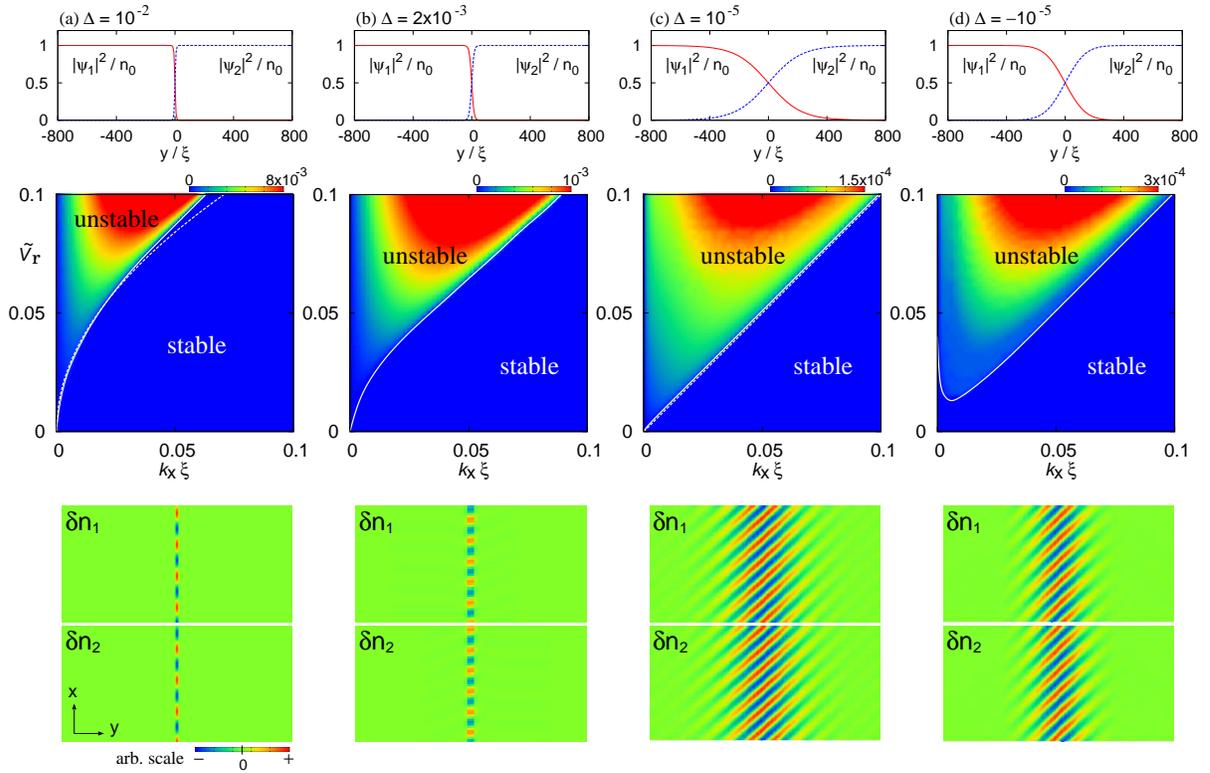}
\caption{
(Color online) Density distributions of the ground states (top panels),
	maximum values of the imaginary part in the Bogoliubov spectra (middle
	panels), and profiles $\delta n_1(x, y)$ and $\delta n_2(x, y)$ of the
	most unstable modes defined in Eq.~(\ref{F}) (bottom panels).
The field gradient $G$ is 0 in (a)--(c) and $\tilde G = m \xi^3 G /
\hbar^2 = 10^{-7}$ in (d).
In the middle panels, the solid lines indicate the boundaries between
dynamically stable and unstable regions and the dashed lines are plots of
Eq.~(\ref{boundary1}) in (a) and Eq.~(\ref{linear}) in (c).
The color bars are normalized by $g n_0$.
In the bottom panels, the relative velocity is $\tilde{V}_{\rm r} = m \xi
V_{\rm r} / \hbar = 0.08$ and the field of view is $1600 \xi \times 800
\xi$.
All the results are independent of $V_1 + V_2$.
}
\label{f:bogo}
\end{figure*}
We consider three types of interfaces:
thin and thick interfaces produced by the intercomponent repulsion and an
interface produced by a potential gradient;
these interfaces are shown in the top
panels of Figs.~\ref{f:bogo} (a), \ref{f:bogo} (c), and \ref{f:bogo} (d),
respectively.
The interface thickness produced by the intercomponent repulsion is
characterized by
\begin{equation}
\Delta = \frac{g_{12}}{g} - 1,
\end{equation}
where $g \equiv g_{11} = g_{22}$.
For $\Delta \ll 1$, the density distribution is approximated
as~\cite{Barankov}
\begin{equation}
n_j(y) = \frac{n_0}{2} \left[ 1 - (-1)^j \tanh \frac{\sqrt{2 \Delta}
y}{\xi} \right],
\end{equation}
where $\xi = \hbar / (m g n_0)^{1/2}$ is the healing length.
The interface thickness is therefore $\simeq \xi \Delta^{-1/2}$.
The potential gradient in Fig.~\ref{f:bogo} (d) has the form $U_1 = -U_2 =
G y$ with $G > 0$ being a constant, which stabilizes the interface at $y =
0$.

To observe KHIs and CSIs in a phase-separated BEC, we perform Bogoliubov
analysis.
Substituting the stationary state~(\ref{Psi}) and a small excitation of
the form,
\begin{equation}
\phi_j = u_{j k_x}(y) e^{i k_x x - i \omega t}
- v_{j k_x}^*(y) e^{-i k_x x + i \omega t},
\end{equation}
into the GP equation and taking the first order of $\phi_j$, we obtain the
Bogoliubov-de Gennes (BdG) equation similar to Eq.~(\ref{bogo2}), in which
$\bm{K}_j$ has only the $x$ component and $n_j$, $u_{j \bm{k}}$, and $v_{j
\bm{k}}$ are replaced by $n_j(y)$, $u_{j k_x}(y)$, and $v_{j k_x}(y)$,
respectively.
The BdG equation is diagonalized numerically.

The middle panels in Fig.~\ref{f:bogo} show the maximum value of the
imaginary part of the Bogoliubov frequency Im $\omega$ for each relative
velocity $V_{\rm r} = V_1 - V_2$ and wave number $k_x$ of the excitation,
where the solid lines divide the dynamically stable and unstable regions.
The stability boundary for the KHI in Eq.~(\ref{KHIrange}) with $\rho_1 =
\rho_2 = m n_0$ has the form
\begin{equation} \label{boundary1}
k_x = \frac{m n_0 V_{\rm r}^2}{2\alpha}.
\end{equation}
The dashed line in the middle panel of Fig.~\ref{f:bogo} (a) plots
Eq.~(\ref{boundary1}) with the interface tension coefficient in
Refs.~\cite{Barankov,Schae} as
\begin{equation} \label{alpha}
\alpha = \frac{\hbar n_0^{3/2}}{\sqrt{2m}} \sqrt{g_{12} - g}.
\end{equation}
The solid and dashed lines agree well for a low relative velocity.
If $G > 0$, the stability boundary deviates from Eq.~(\ref{boundary1}) for
$k \rightarrow 0$~\cite{Takeuchi}.
The dashed line in the middle panel of Fig.~\ref{f:bogo} (c) plots 
\begin{equation} \label{linear}
k_x = \frac{m}{\hbar} V_{\rm r},
\end{equation}
which is the stability boundary for the CSI in Eq.~(\ref{krange}) with $u
= u_{12}$ and $\chi = 0$.
The dashed line almost overlaps with the solid line.
These facts indicate that the KHI is the dominant instability for the
parameters in Fig.~\ref{f:bogo} (a), whereas the CSI dominates in
Fig.~\ref{f:bogo} (c).

The dominant instability changes from the KHI to the CSI when the
interface thickness or the relative velocity increases.
The crossover between the two instabilities can be estimated by equating
Eqs.~(\ref{boundary1}) and (\ref{linear}),
\begin{equation}
\frac{m n_0 V_{\rm r}^2}{2\alpha} \sim \frac{m}{\hbar} V_{\rm r},
\end{equation}
which can be rewritten as
\begin{equation} \label{cross}
\frac{\hbar}{m V_{\rm r}} \sim \frac{\xi}{\sqrt{\Delta}}
\end{equation}
using Eq.~(\ref{alpha}).
For example, for $\Delta = 2 \times 10^{-3}$, the crossover velocity is
$\tilde{V}_{\rm r} = m \xi V_{\rm r} / \hbar \sim 0.04$.
This is consistent with the middle panel of Fig.~\ref{f:bogo} (b), in
which the stability boundary is parabolic for $\tilde{V}_{\rm r} \ll 0.04$
and linear for $\tilde{V}_{\rm r} \gg 0.04$.
Equation (\ref{cross}) indicates that the KHI is the dominant instability
when the interface thickness $\sim \xi \Delta^{-1/2}$ is much
smaller than the characteristic wavelength $\hbar / (m V_{\rm r})$
associated with the relative velocity.
The CSI is dominant in the opposite limit, i.e.,  $\xi \Delta^{-1/2} \gg
\hbar / (m V_{\rm r})$.
It should be noted that the stability boundaries in the middle panels of
Figs.~\ref{f:bogo} (a)-\ref{f:bogo} (c) have the universal form scaled by
$\Delta^{1/2}$.
This is because the healing length $\xi$ is not relevant in the present
problem, and the characteristic length scale for $G = 0$ is only the
interface thickness $\xi \Delta^{-1/2}$.
In fact, plotting the boundaries with respect to $k_x \xi \Delta^{1/2}$
and $\tilde V_{\rm r} \Delta^{1/2}$, we find that they agree very well.

Another difference between the KHI and the CSI is found in the excitation
modes.
The bottom panels in Fig.~\ref{f:bogo} show the change in the density by
excitation of the most unstable mode,
\begin{eqnarray} \label{F}
\delta n_j(x, y) & = & \left| \sqrt{n_j(y)} + c u_{j k_x}(y) e^{i k_x x} -
	c v_{j k_x}^*(y) e^{-i k_x x} \right|^2 \nonumber \\
& & - n_j(y), 
\end{eqnarray}
where $c \ll 1$ is a small constant.
The first term on the right-hand side of Eq.~(\ref{F}) is the density
distribution with a small excitation of the mode.
We see that the excitation modes are localized near the interface.
In the bottom panel of Fig.~\ref{f:bogo} (a), the wavelength of the mode
is larger than the interface thickness, and the excitation of the
mode shifts the interface sinuously.
On the other hand, in the bottom panel of Fig.~\ref{f:bogo} (c), the
wavelength of the mode is smaller than the interface thickness, and
the dynamical instability develops stripes in the interface region.
Interestingly, the stripes are inclined with respect to
the interface, whereas the most unstable wave vector is parallel to the
relative velocity in the uniform system discussed in Sec.~\ref{s:counter}.

Figure~\ref{f:bogo} (d) shows the case of the interface produced by the
field gradient $G > 0$, where the miscible condition $\Delta < 0$ is
satisfied.
The stability boundary in the middle panel of
Fig.~\ref{f:bogo} (d) is similar to that for the immiscible case with the
field gradient (Fig.~2 of Ref.~\cite{Takeuchi}).
The excitation profile in the bottom panel of Fig.~\ref{f:bogo} (d) is
inclined stripes, just as in Fig.~\ref{f:bogo} (c).
If we apply the same field gradient to the system of Fig.~\ref{f:bogo}
(c), we obtain results similar to Fig.~\ref{f:bogo} (d) (data not shown).

\begin{figure}[t]
\includegraphics[width=8.0cm]{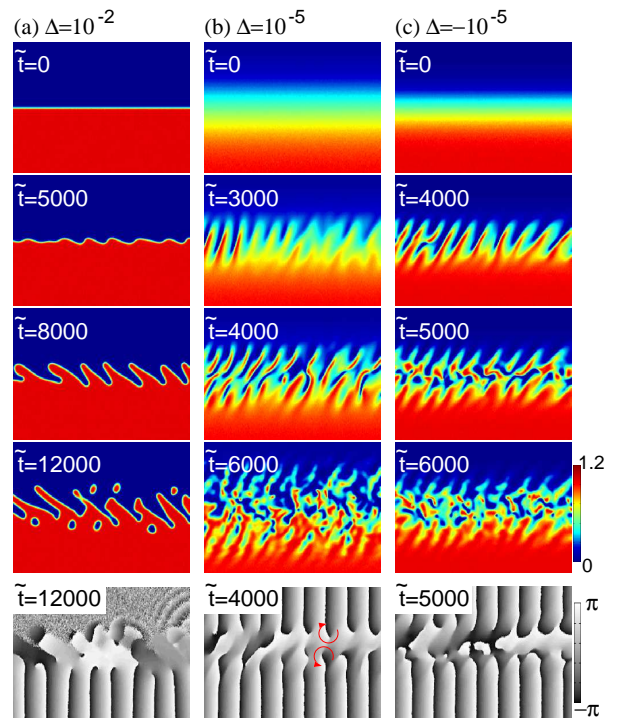}
\caption{
(Color online) Time evolution of the two-component BEC for (a) $\Delta =
 10^{-2}$ with $\tilde G = 0$, (b) $\Delta = 10^{-5}$ with $\tilde G = 0$,
 and (c) $\Delta = -10^{-5}$ with $\tilde G = 10^{-7}$.
The relative velocity is $\tilde{V}_{\rm r} = 0.1$ with $V_1 + V_2 = 0$.
The four upper rows are the density profiles $|\psi_1|^2$ and the bottom
panels are the phase profiles arg $\psi_1$.
The arrows in the bottom panel indicate an example of a vortex pair.
The density is normalized by $g n_0$.
Time $t$ is normalized as $\tilde t = g n_0 t / \hbar$.
The field of view is $1000 \xi \times 750 \xi$.
}
\label{f:2d}
\end{figure}
Figure~\ref{f:2d} demonstrates the time evolution of the immiscible system
for $\Delta = 10^{-2}$ and $10^{-5}$ with no field gradient $G = 0$
and the miscible system for $\Delta = -10^{-5}$ with a field gradient $G
> 0$, where the relative velocity is $\tilde{V}_{\rm r} = 0.1$ with $V_1 +
V_2 = 0$.
The initial state is the stationary state of the GP equation with small
white noise to break the translation symmetry and trigger the dynamical
instability.
The time evolution is obtained by solving the GP equation by the
pseudospectral method.
For $\Delta = 10^{-2}$ [Fig.~\ref{f:2d} (a)], the KHI is dominant,
corresponding to Fig.~\ref{f:bogo} (a).
We see that the interface is modulated due to the KHI ($\tilde t = g n_0 t
/ \hbar = 5000$).
In Figs.~\ref{f:2d} (b) and ~\ref{f:2d} (c), the interface is much
thicker and the CSI is dominant; 
modulation arises in the region where the two components overlap.
The nonlinear dynamics and vortex creation are quite different in the
cases of the KHI and the CSI.
In the former, quantized vortices are formed at the peaks and
troughs of the wave, and enter into each component from the
interface~\cite{Takeuchi}.
In contrast, in the latter, vortex pairs are created in the
interface region (arrows in the bottom panels), which are subsequently
disturbed in a complicated manner.
In all the cases in Fig.~\ref{f:2d}, the total density $|\psi_1|^2 +
|\psi_2|^2$ is almost constant throughout the dynamics.

\section{Instabilities in pancake-shaped systems}
\label{s:quasi}

\subsection{Kelvin-Helmholtz instability}
\label{s:tKHI}

We have considered an ideal flat interface in Sec.~\ref{s:ideal}.
In the following sections, we propose realistic systems for experimental
observation of interfacial instabilities in trapped BECs.

We first consider an axisymmetric harmonic potential given by
\begin{equation} \label{trap}
U_1(\bm{r}) = U_2(\bm{r}) = \frac{1}{2} m [\omega_{\perp}^2 (x^2 + y^2) +
\omega_z^2 z^2],
\end{equation}
where the radial and axial trap frequencies are $\omega_{\perp} = 2 \pi
\times 80$ Hz and $\omega_z = 2 \pi \times 4$ kHz, respectively.
Since $\hbar \omega_z$ is much larger than other characteristic energies,
we reduce the system to two dimensions (2D) in the simulation.
Assuming that the wave function can be written as $\psi(\bm{r}) =
\psi_\perp(x, y) \psi_z(z)$ with $\psi_z(z)$ being the ground state of the
harmonic oscillator, and integrating the GP equation with respect to $z$,
we obtain the 2D GP equation with the effective interaction coefficient
$g_{jj'}^{\rm 2D} = [m \omega_z / (2 \pi \hbar)]^{1/2} g_{jj'}$.
Components 1 and 2 are assumed to be respectively the hyperfine states
$|F, m_F \rangle = |1,0 \rangle$ and $|1, 1 \rangle$ of $^{23}{\rm Na}$
atoms.
The scattering lengths measured in Refs.~\cite{Crub,Black} give $ a_{11}
\simeq 53.4 a_{\rm B}$ and $a_{12}= a_{22} \simeq 54.2 a_{\rm B}$ with
$a_{\rm B}$ being the Bohr radius, which satisfy the condition of phase
separation (\ref{immiscible}).
A strong magnetic field suppresses the spin exchange dynamics, $|0, 0
\rangle |0, 0 \rangle \rightarrow |1, 1 \rangle |1, -1 \rangle$,  
due to the quadratic Zeeman effect~\cite{Miesner}.

\begin{figure}[t]
\includegraphics[width=8.5cm]{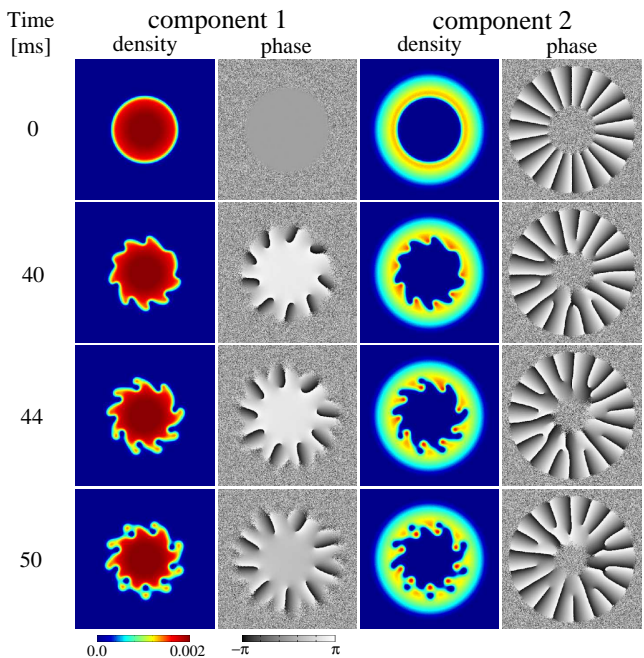}
\caption{
(Color online) Time evolution of the density and phase profiles.
In the initial state, component 2 rotates with vorticity $n_2 = 18$
and component 1 is at rest.
The total number of $^{23}{\rm Na}$ atoms is $N = 1.63 \times 10^6$ with
an equal population in each component.
The field of view is $ 93 \times 93 $ $\mu{\rm m}$.
The unit of density is $N / a^{2}_{\perp}$ with $a_{\perp} = [\hbar /
(m \omega_{\perp})]^{1/2}$.
}
\label{f:2dna}
\end{figure}
The initial state is prepared as follows.
We first calculate the ground state $|\psi_j|$ with a centrifugal
potential $\hbar^2 n_j^2 / [2 m (x^2 + y^2)]$ by the imaginary-time
propagation of the GP equation.
We then give the phases as
\begin{equation}
\psi_j = e^{in_{j}\theta} |\psi_j| \qquad (j = 1,2),
\end{equation}
where the integer $n_j$ is the vorticity of the $j$th component and $\theta
= {\rm arg} (x + i y)$.
This procedure is efficient for preparing an axisymmetric rotating
stationary state.
We thus obtain a stationary state with shear flow at the interface, as
shown in the top panels in Fig.~\ref{f:2dna}, where $n_1 = 0$ and $n_2 =
18$.
Because $a_{22} > a_{11}$ and the centrifugal force on component 2,
component 1 occupies the center and component 2 surrounds it.
The presence of component 1 stabilizes the vortices in component 2.
We add small numerical noise to the initial state to break the
axisymmetry.
The two components are assumed to have the same number of atoms in the
following analysis.

Figure~\ref{f:2dna} shows the subsequent time evolution.
At $t = 40$ ms, a wavy pattern develops at the interface due to the
KHI~\cite{Takeuchi}, which has approximately nine-fold symmetry.
The wave at the interface then grows ($t = 44$ ms) and quantized vortices
are released from the peaks and the troughs of the wave into both
components ($t = 50$ ms).
This interface behavior is similar to that in the flat interface
[Fig.~\ref{f:2d} (a)].
When the relative velocity at the interface is below a critical velocity
($n_2 < 12$), symmetry-breaking dynamics does not occur, whereas the flat
interface is always unstable for a nonzero relative velocity [see
Eq.~(\ref{KHIrange})].
This indicates that in the trapped system the effective force is exerted
on the interface [see Eq.~(\ref{Vcr})], which originates from the trap
potential and the centrifugal force.

\begin{figure}[t]
\includegraphics[width=8.0cm]{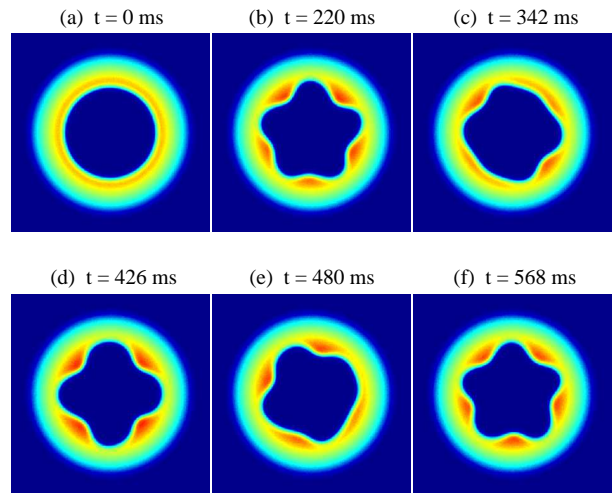}
\caption{
(Color online) Time evolution of the density profile of component 2 for
$n_2 = 12$.
Other parameters are the same as those in Fig.~\ref{f:2dna}.
}
\label{f:2dna2}
\end{figure}
Figure~\ref{f:2dna2} shows the time evolution of component 2 for a
relative velocity just above the critical value ($n_2 = 12$).
A five-fold pattern first emerges [Fig.~\ref{f:2dna2} (b)], which
changes to a four-fold pattern [Fig.~\ref{f:2dna2} (d)] and then returns
to a five-fold pattern [Fig.~\ref{f:2dna2} (f)].
In contrast to the case of $n_2 = 18$ shown in Fig.~\ref{f:2dna}, the
deformation of the interface is moderate and no quantized vortices are
generated throughout the dynamics.
Estimating the unstable wavelengths in Figs.~\ref{f:2dna}
and \ref{f:2dna2} from the analytic expression (\ref{mu}) is difficult
because of an ambiguity in the interface tension coefficient $\alpha$ in
inhomogeneous systems with a curved interface.

\subsection{Bogoliubov analysis}

We perform Bogoliubov analysis of a quasi-2D trapped system.
We write the wave function as
\begin{equation} \label{psi2d}
\psi_j(r, \theta) = \left[ f_j(r) + \phi_j(r, \theta) \right] e^{i n_j
	\theta} e^{-i \mu_j t / \hbar},
\end{equation}
where $f_j(r) e^{i n_j \theta}$ is the stationary state, $\phi_j$ is a
small excitation, $\mu_j$ is the chemical potential, and $r = (x^2 +
y^2)^{1/2}$.
We expand the small excitation $\phi_j$ as
\begin{equation} \label{phi2d}
\phi_j(r, \theta) = u_{jL}(r) e^{i L \theta - i \omega t} - v_{jL}^*(r)
	e^{-i L \theta + i \omega t},
\end{equation}
where $L$ is an integer and the excitation has $L$-fold symmetry.
Substituting Eqs.(\ref{psi2d}) and (\ref{phi2d}) into the GP equation and
taking the first order of $\phi_j$, we obtain the BdG equation ($j \neq
j'$),
\begin{widetext}
\begin{subequations} \label{BdG2d}
\begin{eqnarray}
\left\{ -\frac{\hbar^2}{2m} \left[ \frac{\partial^2}{\partial r^2} +
	\frac{1}{r} \frac{\partial}{\partial r} - \frac{(n_j + L)^2}{r^2}
	\right] + U_j - \mu_j + 2 g_{jj}^{\rm 2D} f_j^2 + g_{jj'}^{\rm 2D}
	f_{j'}^2 \right\} u_{jL} 
- g_{jj}^{\rm 2D} f_j^2 v_{jL} + g_{jj'}^{\rm 2D} f_j f_{j'} (u_{j'L}
	- v_{j'L}) & = & \hbar \omega u_{jL}, \nonumber \\
\\
\left\{ -\frac{\hbar^2}{2m} \left[ \frac{\partial^2}{\partial r^2} +
	\frac{1}{r} \frac{\partial}{\partial r} - \frac{(n_j - L)^2}{r^2}
	\right] + U_j - \mu_j + 2 g_{jj}^{\rm 2D} f_j^2 + g_{jj'}^{\rm 2D}
	f_{j'}^2 \right\} v_{jL} 
- g_{jj}^{\rm 2D} f_j^2 u_{jL} - g_{jj'}^{\rm 2D} f_j f_{j'} (u_{j'L}
	- v_{j'L}) & = & -\hbar \omega v_{jL}. \nonumber \\
\end{eqnarray}
\end{subequations}
\end{widetext}
We numerically diagonalize Eq.~(\ref{BdG2d}).

\begin{figure}[t]
\includegraphics[width=8.5cm]{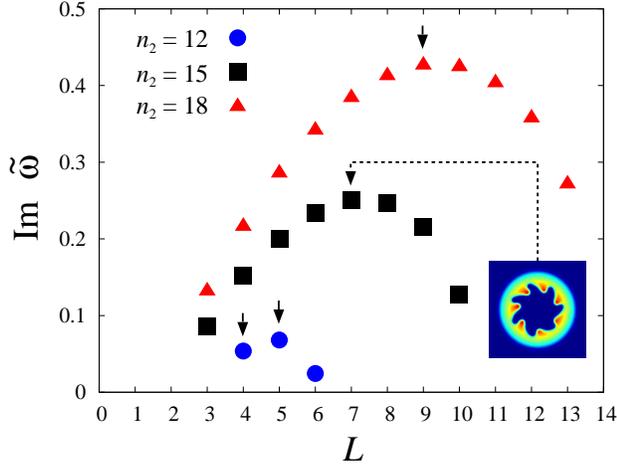}
\caption{
(Color online) Imaginary part of Bogoliubov frequency $\tilde\omega =
	\omega / \omega_\perp$ for $n_1 = 0$ and $n_2 = 12$ (circles), $n_2 =
	15$ (squares), and $n_2 = 18$ (triangles).
The excitation mode has $L$-fold symmetry.
The total number of atoms is $N = 1.63 \times 10^6$.
The conditions for triangles and circles are the same as those in
Figs.~\ref{f:2dna} and \ref{f:2dna2}, respectively.
The solid arrows indicate the modes relevant to the patterns in
Figs.~\ref{f:2dna} and \ref{f:2dna2}.
The inset shows a snapshot of the density profile of component 2 obtained
by time evolution for $n_2 = 15$.
}
\label{f:KHI2Dbogo}
\end{figure}
Figure~\ref{f:KHI2Dbogo} shows the imaginary part of the excitation
frequency for the same conditions as those in Fig.~\ref{f:2dna}
($n_2 = 18$, triangles) and Fig.~\ref{f:2dna2} ($n_2 = 12$, circles).
For $n_2 = 18$, the imaginary part is a maximum for $L = 9$, which is
consistent with the nine-fold pattern in Fig.~\ref{f:2dna}.
The emergence of the four and five-fold patterns in Fig.~\ref{f:2dna2} is
understood from the fact that the corresponding imaginary parts are close
to each other (the circles indicated by the arrows in
Fig.~\ref{f:KHI2Dbogo}).
In fact, we numerically confirmed that for $n_2 = 18$, a nine-fold pattern
or a 10-fold pattern emerges depending on the initial random noise.
For $n_2 < 12$, no imaginary part appears.
For the condition in Fig.~\ref{f:KHI2Dbogo}, only the lowest mode becomes
dynamically unstable for each $L$.

\subsection{Counter-superflow instability}
\label{s:mix2D}

We next consider the CSI in the pancake-shaped system.
We use the hyperfine states $|1, 1 \rangle$ and $|1, 0 \rangle$ of
$^{87}{\rm Rb}$ atoms for components 1 and 2.
According to Ref.~\cite{Kempen}, the scattering lengths are $a_{11}=
a_{12} = 100.4 a_{\rm B}$ and $a_{22}=100.9 a_{\rm B}$, which satisfy the
miscible condition~(\ref{miscible}).
We employ the harmonic potential plus a central optical plug given by
\begin{eqnarray} \label{plug}
U_1(\bm{r}) = U_2(\bm{r}) & = & \frac{1}{2} m [\omega_{\perp}^2 (x^2 +
	y^2) + \omega_z^2 z^2] 
\nonumber \\
& & + \hbar \omega_{\perp} \alpha e^{-\beta (x^2 + y^2) /	a_{\perp}^2},
\end{eqnarray}
where $\alpha$ and $\beta$ are dimensionless parameters respectively
characterizing the strength and the width of the optical plug, and
$a_{\perp} = [\hbar / (m \omega_{\perp})]^{1/2}$.
The same trap frequencies as those in Sec.~\ref{s:tKHI} are used and hence
the calculation is performed in 2D.
We choose the parameters of the optical plug to be $\alpha = 1000$ and
$\beta = 0.01$.
The optical plug is necessary for $n_1 \neq 0$ and $n_2 \neq 0$ to
eliminate undesirable instability.
If the optical plug is absent, one component flows into the vortex core of
the other component and the axisymmetry is broken before the CSI emerges.

\begin{figure}[t]
\includegraphics[width=8.5cm]{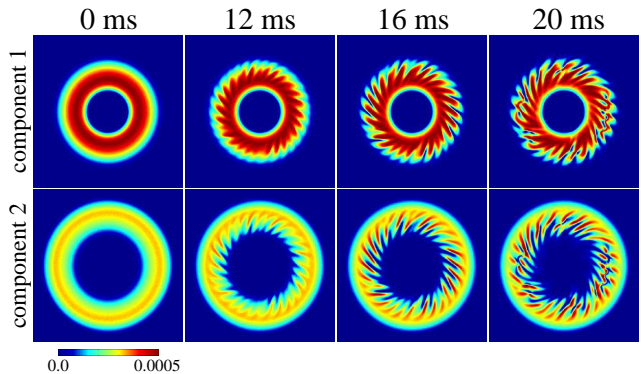}
\caption{
(Color online) Time evolution of the density profiles for counter-rotating
miscible condensates.
The vorticities of the initial state are $n_1 = -20$ and $n_2 = 30$.
The total number of $^{87}{\rm Rb}$ atoms is $N = 3.19 \times 10^6$ with
an equal population in each component. 
The field of view is $84 \times 84 $ $\mu{\rm m}$.
}
\label{f:mix}
\end{figure}
Figure~\ref{f:mix} shows the time evolution of counterrotating condensates
with initial vorticities $n_1 = -20$ and $n_2 = 30$.
Despite the miscible condition, the two condensates are weakly separated
($t = 0$ ms of Fig.~\ref{f:mix}) because of the difference in the
centrifugal force.
As time develops, a saw-toothed pattern emerges at the interface, as shown
in Fig.~\ref{f:mix}.
This behavior at the interface is similar to that in Figs.~\ref{f:2d} (b)
and \ref{f:2d} (c), indicating that the symmetry-breaking dynamics is
driven by the CSI.
The relation between the CSI and quantum turbulence has recently been
studied~\cite{Ishino}.

\section{Instabilities in cigar-shaped systems}
\label{s:3D}

\subsection{Kelvin--Helmholtz instability}
\label{s:type1}

We propose two systems for observing the KHI in a cigar-shaped geometry.

The first one is similar to that in Sec.~\ref{s:tKHI}; non-rotating
component 1 is surrounded by rotating component 2.
We use a harmonic potential in Eq.~(\ref{trap}) with $\omega_{\perp} =
2\pi \times 40$ Hz and $\omega_z = 2 \pi \times 8$ Hz.
Components 1 and 2 are respectively the hyperfine states $|2, 1 \rangle$
and $|1, -1 \rangle$ of $^{87}{\rm Rb}$~\cite{Note}.
The ratio between the scattering lengths of these states is
$a_{11}:a_{12}:a_{22}= 0.97:1.0:1.03$ with their average being $5.5$
nm~\cite{Matthews,Hall}, which satisfy the immiscible condition
(\ref{immiscible}).

\begin{figure}[t]
\includegraphics[width=8.5cm]{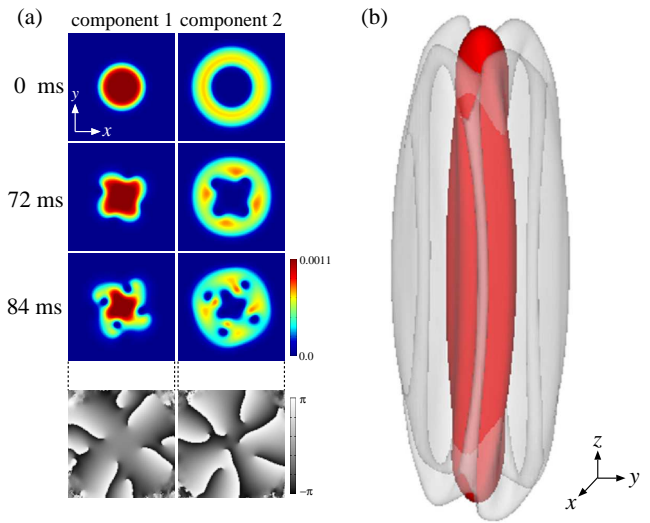}
\caption{
(Color online) (a) Time evolution of the cross-section densities
$|\psi_1(x, y, 0)|^2$ (left panels) and $|\psi_2(x, y, 0)|^2$ (right
panels)	for $n_1 = 0$ and $n_2 = 7$.
The field of view is $20 \times 20$ $\mu{\rm m}$.
The total number of $^{87}{\rm Rb}$ atoms is $N = 1.75 \times 10^5$ with an
equal population in each component.
The bottom panels show the phase profiles at $t = 84$ ms.
(b) Isodensity surfaces of component 1 (inside) and component 2 (outside)
at $t = 84$ ms.
For clarity, only the surface of component 2 is transparent.
}
\label{f:ring}
\end{figure}
We solve the GP equation by performing a 3D numerical simulation.
Figure~\ref{f:ring} (a) shows the time evolution of the cross-section
densities $|\psi_{1}(x,y,0)|^2$ and $|\psi_{2}(x,y,0)|^2$ for the initial
vorticities $n_1 = 0$ and $n_2 = 7$.
The initial state has an axisymmetric interface with a relative velocity.
At $t = 72$ ms, a wavy pattern with four-fold symmetry develops at the
interface due to the KHI.
Vortex lines are then released from the interface into both components
at $t = 84$ ms as shown in Figs.~\ref{f:ring} (a) and \ref{f:ring} (b).
For a smaller relative velocity at the interface (e.g., $n_1 = 0$ and $n_2
= 1$), symmetry breaking dynamics does not occur, indicating the existence
of a critical velocity, as in the case of Sec.~\ref{s:tKHI}.

\begin{figure}[t]
\includegraphics[width=8.5cm]{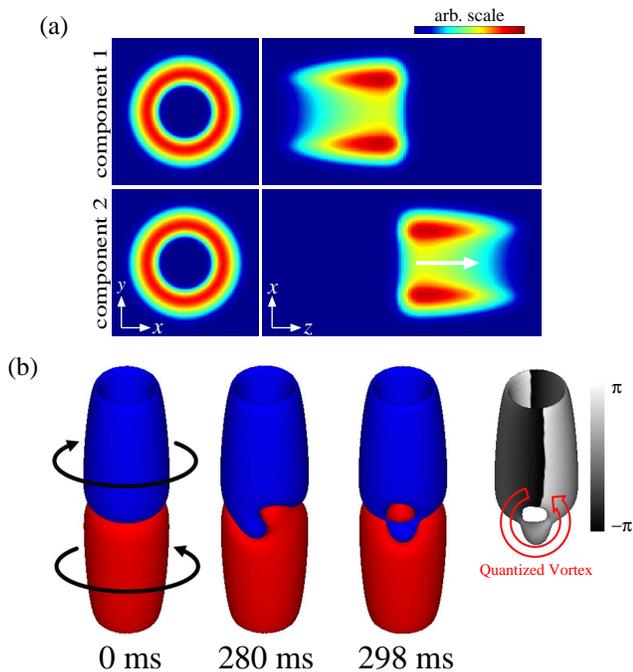}
\caption{
(Color online) (a) Column densities $\int dz |\psi_j|^2$
(left panels, $40 \times 40 $ $\mu{\rm m}$) and $\int dy |\psi_j|^2$ 
(right panels, $40 \times 132 $ $\mu{\rm m}$) of the initial state for
	$n_1 = 2$ and $n_2 = -2$.
The white arrow indicates the direction of the Stern-Gerlach force
produced by the field gradient of $dB/dz = -17.3$ mG/cm.
(b) Time evolution of the isodensity surfaces of components 1 (lower) and
2 (upper).
The solid arrows at $t = 0$ ms indicate the directions of the rotation.
The phase profile of component 2 at $t = 298$ ms is shown in the rightmost
image.
The total number of $^{23}{\rm Na}$ atoms is $N = 6.6 \times 10^5$ with an
equal population in each component.
}
\label{f:cigar}
\end{figure}
The second configuration for observing the KHI in a cigar-shaped trap is
shown in Fig.~\ref{f:cigar} (a), where components 1 and 2 are located in
the $z \lesssim 0$ and $z \gtrsim 0$ regions and their interface lies on
$z \simeq 0$.
We use the same hyperfine states of $^{23}{\rm Na}$ as those used in
Sec.~\ref{s:tKHI}.
The potential in Eq.~(\ref{plug}) is used with $\omega_{\perp} =
2\pi \times 40$ Hz, $\omega_z = 2 \pi \times 8$ Hz, $\alpha = 50$, and
$\beta = 0.4$.
The optical plug is applied to prevent one component flowing into the
vortex core of the other component.
We apply a magnetic field gradient of $dB/dz = -17.3$ mG/cm, which pushes
component 2 ($|1, 1 \rangle$) in the $+z$ direction.
This force controls the strength of the KHI.

Figure~\ref{f:cigar} (b) depicts the time evolution of the system with
initial vorticities of $n_1 = 2$ and $n_2 = -2$.
The relative velocity at the interface induces the KHI, which breaks the
axisymmetry and deforms the interface [$280$ ms of Fig.~\ref{f:cigar}
(b)].
A quantized vortex is then released from the interface into each component
at $t \simeq 298$ ms.
The vortex lines lie in the radial direction;
in contrast, the vortex lines lie along the $z$ direction in
Fig.~\ref{f:ring}.

\subsection{Counter-superflow instability}
\label{s:mix3D}

As shown in Sec.~\ref{s:ideal}, the interfacial instability is dominated
by the CSI when the overlap between the two components is large.
We study this situation for a cigar-shaped system.
The two components are the same hyperfine states of $^{87}{\rm Rb}$ as
those used in Sec.~\ref{s:mix2D}, and the potential is the same as that in
Sec.~\ref{s:type1}.

\begin{figure}[t]
\includegraphics[width=7.5cm]{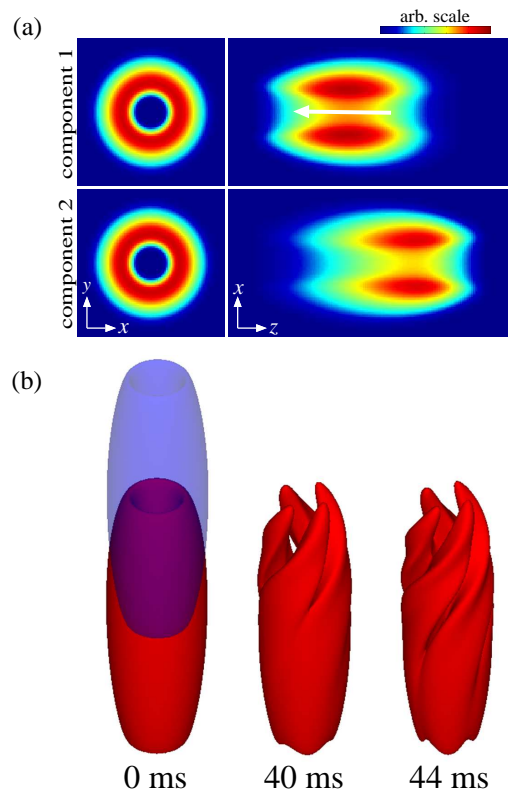}
\caption{
(Color online) (a) Column densities $\int dz |\psi_j|^2$ 
(left panels, $27 \times 27 $ $\mu{\rm m}$) and $\int dy |\psi_j|^2$ 
(right panels, $27 \times 118 $ $\mu{\rm m}$) of the initial state for
$n_1 = 3$ and $n_2 = -3$.
The white arrow shows the direction of the Stern--Gerlach force produced
by a field gradient of $dB / dz = 0.34$ mG/cm.
(b) Time evolution of the isodensity surface of component 1.
At $t = 0$, component 2 is also shown as the transparent surface.
The total number of $^{87}{\rm Rb}$ atoms is $N = 1.3 \times 10^6$ with an
equal population in each component.
}
\label{f:mix3d}
\end{figure}
Figure~\ref{f:mix3d} shows the time evolution of the system for initial
vorticities of $n_1 = 3$ and $n_2 = -3$.
Since the scattering lengths satisfy the miscible condition and the field
gradient is small, the two components in the initial stationary state
widely overlap with each other [Fig.~\ref{f:mix3d} (a)].
The time evolution in Fig.~\ref{f:mix3d} (b) is similar to those in
Figs.~\ref{f:2d} (b) and \ref{f:2d} (c) in the sense that stripes of
the two components are formed where they overlap.
The spiral pattern in Fig.~\ref{f:mix3d} (b) ($t \simeq 40$ ms)
corresponds to the inclination of the stripes in Figs.~\ref{f:bogo}
(d) and \ref{f:2d} (c).

\section{Conclusions}
\label{s:conclusion}

We have studied the dynamical instabilities at an interface of a
two-component BEC with a relative velocity between the two components.
When the two components are strongly segregated and the interface
thickness is negligible, a KHI is generated at the interface.
On the other hand, when the interface thickness is much larger than the
unstable wavelength, a CSI dominates over the KHI.
We have proposed realistic experimental systems of $^{23}{\rm Na}$ and
$^{87}{\rm Rb}$ BECs in pancake-shaped and cigar-shaped traps.
In a pancake-shaped trap, a KHI can be observed at the interface of two
components separated inside and outside, and rotating with different
vorticities (Figs.~\ref{f:2dna} and \ref{f:2dna2}).
The patterns produced by the KHI are consistent with the Bogoliubov
analysis (Fig.~\ref{f:KHI2Dbogo}).
Using miscible two components, a CSI can also be realized
(Fig.~\ref{f:mix}).
For a cigar-shaped trap, we proposed two configurations for observing the
KHI: the two components separate radially (Fig.~\ref{f:ring}) or in the
$z$ direction (Fig.~\ref{f:cigar}).
The vortex lines released from the interface lie along the radial
direction in the former, whereas they lie along the $z$ direction in the
latter.
The interface thickness can be controlled using the miscible condensates
with a magnetic field gradient in the $z$ direction (Fig.~\ref{f:mix3d}).

The initial states used in Secs.~\ref{s:quasi} and \ref{s:3D} are multiply
quantized vortices with different vorticities in the two components.
Such states can be prepared using Laguerre--Gaussian
beams~\cite{Andersen} or topological phase
imprinting~\cite{Leanhardt}.
Multiply quantized vortex states with large vorticities can in principle
be generated by repeated application of these methods~\cite{Mottonen,Xu}.

The dynamics caused by the KHI and the CSI in quantum fluids are quite
different from those in classical fluids, since vortices are quantized and
there is no viscosity.
Moreover, in the present system, the miscibility of the two components can
be dynamically controlled by Feshbach resonance~\cite{Papp,Tojo}.
These features of BECs renew our interest in fluid
instabilities~\cite{Sasaki,Gautam,Karman,Bezett} and may provide new physical
insights into fluid dynamics.

\begin{acknowledgments}  
This work was supported by KAKENHI from JSPS and MEXT (Nos.\ 199748,
 17071005, 17071008, 20540388, 21340104, 21740267, and 22340116).
\end{acknowledgments}

\end{document}